\documentstyle[12pt,epsfig]{article}
\newcommand{\ba}{\begin{eqnarray}}
\newcommand{\ea}{\end{eqnarray}}

\begin{document}

\author{Rama CONT\\
Centre de Math\'ematiques Appliqu\'ees\\
  Ecole Polytechnique \\
F-91128 Palaiseau, France\\
E-mail: Rama.Cont@polytechnique.fr}

\title{Modeling term structure dynamics:\vskip 0.5cm
an infinite dimensional approach
\footnote{Part of this work was done while the author
was a research fellow at the Institute of Theoretical Physics, 
Swiss Federal Institute of Technology, Lausanne (EPFL). I am indebted to Nicole El Karoui for introducing me
to this subject and sharing with me her experience in interest rate modeling. 
I also thank Toufik Abboud, Jean-Philippe Bouchaud, Robert Dalang, St\'ephane Denise, Olivier
Lev\^eque, Carl Mueller, Francesco Russo and Agn\`es Sulem for useful indications and encouragement.
 All remaining errors are mine. }
}
\date{\ }
\maketitle

\vfill
Keywords: stochastic PDE, cylindrical brownian motion, stochastic evolution equations,
random fields, term structure of interest rates, forward rates, 
parabolic equations, random strings, multifactor models.
\newpage

\begin{abstract}
We present a family of models for term structure dynamics 
in an attempt to describe 
several statistical features 
observed in empirical studies of forward rate curves
by decomposing the deformations of the term structure into
the variations of the short rate, the long rate and 
the fluctuations of the curve around its average shape.
This fluctuation is then described as a solution of a stochastic evolution
equation in an infinite dimensional space.
In the case where deformations are local in maturity, this equation reduces to a stochastic PDE, of which we give the simplest example.
We discuss the properties of the solutions and show that they capture
in a parsimonious manner
the essential features of yield curve dynamics: imperfect correlation between
maturities, mean reversion of interest rates and the structure of
principal components of term structure deformations.
Finally, we discuss calibration issues and show that
the model parameters have a natural interpretation in terms of empirically
observed quantities.
\vskip 1cm
Nous proposons une description de la d\'eformation de la structure par terme
des taux d'int\'er\^et en termes d'une \'equation aux d\'eriv\'ees partielles
 stochastique. En \'etudiant en d\'etail le cas lin\'eaire, nous montrons
qu'une telle formulation  rend compte des observations empiriques sur la dynamique de la
courbe des taux forwards avec un nombre restreint de param\`etres. Nous
discutons en d\'etail le r\^ole des hypoth\`eses math\'ematiques et leur impact
sur la dynamique des taux d'int\'er\^et.
\end{abstract}

\newpage

\tableofcontents
\newpage

\section{Introduction}

\subsection{Motivations for term structure modeling}

There are two very different motivations for term structure modeling.
The first is concerned with the pricing of interest rate derivative
securities. In this context, which has been the principal motivation
behind term structure models in the mathematical finance literature 
\cite{cir,hjm,vasicek},
the main concern has been the development of ``coherent'' --in the sense of
arbitrage-free -- 
pricing criteria for securities whose payoffs depend on movements of interest rates. 

The second motivation, which could be labeled as ``econometric'',
is  the statistical decription of the movements of real interest rates.
In contrast to the  preceding approach where the  emphasis is on 
cross-sectional
coherence of prices given by the model, here the emphasis is on describing
and reproducing as closely as possible  the time evolution of interest rates
from a statistical point of view. Such an approach is useful if one is interested in simulating scenarios, calculating Value-at-Risk of fixed-income positions
but also from a theoretical point of view, to gain a better understanding 
of interest rate fluctations and their relations to other economic variables..

From a mathematical point of view, the first approach corresponds to modeling
the dynamics of interest rates under a {\it risk-neutral} (or risk-adjusted) measure, while the second approach corresponds to  the modeling of  real-world
term structure dynamics.

Given the complexity of the behavior of the yield curve, there is a conflict 
between these two criteria  which turn out to be difficult to satisfy
in the same model.
Most of the existing work  in the mathematical finance literature on interest
rate modeling has adopted  the first approach, namely the description of yield curve dynamics under a ``risk-adjusted'' measure.  The reason for this trend
is not difficult to understand: contrarily to a stock market, in a bond market 
the future values of many securities -- namely the zero-coupon bonds -- are known with certainty. The observability of the contemporaneous prices of these bonds then
makes it possible to calibrate a model for the risk-adjusted dynamics of interest rates directly to the observed bond prices. However, as correctly
pointed out by various authors (see e.g. \cite{pliska}, pp. 201-204),
once the risk-adjusted dynamics has been calibrated it is not obvious
that such a model will tell us anything useful about the real-world
dynamics of interest rates. 

In fact, it turns to be difficult  to account for a 
series of empirical facts observed in studies on the
term structure of interest rates \cite{string} in the framework of standard term structure
models of the first type. 
On one hand, it seems that the constraints implied by the absence
of arbitrage in these models are
 so strong that the latter is obtained at the detriment of a correct
representation of the dynamics of the yield curve \cite{risk,elk}.
On the other hand, some stylized empirical facts about
term structure deformations seem to have no
theoretical counterpart in classical arbitrage-free models.

\subsection{Continuous models for discrete observations}

Empirical term structure data consist of time series of  interest
rates of various maturities; for example, in the Eurodollar
market interest rates of about 40 different maturities can be obtained
on a daily basis. Instead of modeling the data as  a  sequence of
40-component random vectors, theoretical models tend to
represent the interest rate curve as a function of a {\it continuous}
maturity variable $\theta$ evolving with a 
 continuous time parameter $t$. These models, which offer greater
analytical
tractability, 
correspond to the ideal case where fixed-income 
instruments  for continuum
of maturities are traded in a continuous-time market.
Continuous models for interest rate dynamics can be divided into
two classes.

The first class starts by postulating a stochastic model for
the short term interest rate: typically, the short rate
is considered to follow a stochastic differential equation. 
The model then specifies a rule for constructing the yield curve
from the short rate by an additional set of variables representing
the market price(s) of risk.
This class includes the Vasicek model \cite{vasicek},
 the Cox-Ingersoll-Ross model \cite{cir} and other variants of
these models. Short-rate based models present the advantage of analytical
tractability. However, they reduce the dynamics of the whole yield curve to 
the movement of one it its endpoints, which makes these models rather rigid.
A more serious objection to short-rate based models is that
 empirical studies tend to reject the hypothesis that the short rate 
is Markovian, let alone a diffusion \cite{aitsahalia}.

The second class of models, initiated by Ho \& Lee \cite{holee}
and extended by Heath, Jarrow \& Morton \cite{hjm}, take the initial
term structure and the forward rate volatilities as inputs, model
the evolution of  forward rates as a family of scalar diffusion process
indexed by the maturity date $T$:
\ba\label{hjm.eq}
df_{\sc HJM}(t,T) &=& \alpha(t,T) dt + \sigma(t,T) dW_t
\ea 
The forward rate curve is therefore a continuum of diffusion processes
driven by a {\it finite} number of  noise sources; it is therefore intuitively clear
that if no constraint is imposed on the drift and diffusion coefficients
such a model will present   obvious arbitrage opportunities.
It was shown in \cite{hjm} that,  when the noise source is a finite dimensional
Wiener process, then the absence of an arbitrage
strategy involving bonds of all maturities imposes the following relation between
the drift and the diffusion process of the forward rates:
\ba \label{lien}
\alpha(t,T) = \sigma(t,T) \int_{t}^{T} \sigma(t,u) du + \sigma(t,T) \gamma(t)
\ea
where $\gamma(t)$ is some predictable process independent of the time to maturity $\theta= T-t$.
Underlying this result are the hypotheses that:

- the number of sources of randomness is finite (in practice, much smaller than the number
of maturities)

- arbitrage opportunities involve any number of maturities.

The first hypothesis is a modeling choice which means that
for representing the fluctuations of the yield curve, 
a random vector with around 40 components,
we use a model where the number of components $N$ is infinite (in fact a continuum)
but the number
$k$ of sources of randomness is kept finite.
The well-foundedness of this choice is not obvious a priori
and justified  by observing that Principal Component Analysis (see below)
of yield curve movements reveal that around 95\% 
of the variance of bond yields is explained by 3 factors.
As we shall see below, such an observation does not imply that properties
{\it other} than the covariance structure will be explained by the same
small number of factors.
Moreover, the relation between drift and volatility (Eq.\ref{lien} )
which is fundamental in HJM models is verified if the sources
of randomness defining yield curve dynamics are correctly specified.
What if one neglects a source of randomness when specifiying the model?
This amounts to ``projecting'' the model on the space spanned by the
first few factors or integrating the drift and volatility functions in
Eq.(\ref{hjm.eq}) with respect to the left-out variable. 
In this case  relation (\ref{lien}) has no reason
to hold anymore between the modified drift and volatility.
In other words, the HJM relation (\ref{hjm.eq}) is not robust to
misspecification of the number of sources of randomness.

An alternative approach, which we propose
here, is to consider the limit where $k$ is {\it also} infinite, thus restoring
more flexibility to the forward rate curve and at the same time reducing the
possibility of easy arbitrage \footnote{On arbitrage in infinite factor models, see \cite{douady}.}.
At the same time, we shall take into account the relative segmentation of the fixed
income market into maturity-specific markets to introduce the notion of 
{\it local deformation} and examine its consequences, as explained below.

\subsection{The forward rate curve as an infinite-dimensional process}

In  many 
mathematical models of  the yield curve, the term structure of interest rates
is often parametrized through the {\it forward rate} curve.
Let $B(t,T)$ be the price at time $T$ of a (default-free) zero-coupon
bond with maturity date $T$.
The instantaneous forward rates  $f(t,\theta)$ are related to bond prices by
\ba
\forall t \in [0,T], B(t,T) &=& \exp( -\int_{0}^{T-t} f(t,u) \quad du )
\ea
The forward rate term structure at time $t$ is therefore a function
of a continuous variable $\theta$, representing the time to maturity.
As such, the forward rate curve $f(t,.)$ naturally lives
in an infinite-dimensional space of functions and, as remarked by several authors
\cite{duffie,kennedy} there is no reason {\it a priori} to believe that
its random movements will be restricted  to a finite-dimensional subspace or manifold.
There are two main reasons why the existing theory has 
almost exclusively focused on models with a finite number
of state variables: the first reason is analytical tractability, the second is
that principal component analysis of term structure movements seems to suggest
that a few factors are sufficient to explain the covariance structure of interest rates \cite{string,litterman}.
As pointed out above, making a large (formally infinite) number of forward rates dependent on
 a small  number of factors restricts considerable the various configurations
of term structure movements, creating a conflict between tractability and a faithful
representation of empirical observations.

Note however that, as pointed out by \cite{musiela}, even with a finite-dimensional
noise source in a HJM approach, one cannot ignore the infinite-dimensional
character of the forward  curve process. This means that, in most such
models, the smallest
family of forward rates which is Markovian will be an infinite family.
However, HJM models endow this infinite family of forward rates with only
finitely many degrees of freedom, placing considerable restriction on the
type of dynamics it can follow.
The difference in the present approach
is not only to consider the yield curve as an infinite-dimensional process
but also as a process with infinitely many degrees of freedom.
\subsection{Relation to previous research}

Although the intrinsically infinite-dimensional character of continuous term structure models
has been remarked by many authors, either it has been often
dismissed as being an unrealistic working assumption
 in yield curve modeling because of the mathematical complexity involved, leading to multifactor models of the term structure
(Duffie \& Kan \cite{duffie}, El Karoui {\it et al.} \cite{elk,lacoste,emv}). 
 
Bj\"ork {\it et al.}\cite{bjork} proposed a mathematical framework 
for arbitrage pricing of interest-rate derivatives taking into account
the infinite-dimensional nature of the yield curve process.
However in their representation, following the  HJM \cite{hjm} approach,
this infinite-dimensional process is driven by a {\it finite}-dimensional
source of randomness.

An important step was taken by Musiela's \cite{musiela} representation of the forward rate as a stochastic process
taking its values in a space of functions.
Musiela reformulated the HJM equations in terms of a first-order 
stochastic partial differential equation in the time and maturity variables.
Using this same representation, Goldys \& Musiela subsequently derived
 an infinite-dimensional
version of the Black-Scholes PDE for a swaption \cite{goldys}.

Another direction which has been developed is the representation of the
term structure as a {\it random field} with two parameters, time and maturity.
First introduced by Kennedy \cite{kennedy}, this approach allows to
incorporate 
 an infinite number (or a continuum) 
of sources of randomness in the evolution of the term
structure. Kennedy\cite{kennedy} considers random fields derived from
the Brownian sheet, more general random fields are considered by Goldstein
\cite{goldstein}, who considers as an example random fields which solve a
second-order stochastic PDE.
Building on these examples, 
Santa Clara \& Sornette \cite{santaclaus} consider forward rate
models in which the forward rate process is driven by a two-parameter
noise process (``stochastic string shock''), again pointing out the relation
with stochastic PDEs.

Random field models can also be represented as multifactor models
with an infinite number of factors.
A generalization of the HJM equations to
 Gaussian infinite factor models was  studied
by Douady \cite{douady}. Instead of a random field representation,
Douady uses an infinite-dimensional representation of the yield curve
and introduces a cylindrical Brownian motion as a source of randomness.

An approach which unifies the infinite-dimensional character of the
yield curve process and the possibly large number of possible
sources of randomess
was  developed by Bouchaud {\it et al.}\cite{string,risk}
based on an empirical study of deformations of the Eurodollar term
structure between 1992 ad 1996. 
Bouchaud {\it et al.} proposed the idea of representing the term
structure as a randomly vibrating curve governed by a stochastic partial
differential equation containing a {\it second order} derivative
with respect to maturity and presented some empirical evidence in favor
of such models.

Our objective in this paper is to present a model of term structure movements
which is both analytically tractable, preserves the infinite-dimensional character
of the forward rate curve and reproduces some stylized empirical observations
with a small number of parameters. Our approach  accounts in a natural manner for 
the fact that only a small number of factors seem to govern the covariance structure
of term structure movements, without imposing any {\it ad hoc} unobservable 
state variable in the model.

More precisely, we will try to demonstrate that specifying the term structure process
as a dynamic process with an infinite number of degrees of freedom and eventually proceeding to a finite dimensional
approximation {\it afterwards} is a more robust modeling procedure than restricting
the number of degrees of freedom in the definition of the model, at the risk
of misspecifying the factors.

\subsection{Outline}

The paper is structured as follows. In section \ref{empirical}
we recall some important empirical observations about term structure deformations. Based on these observations, we discuss in section \ref{modeling}
what  ingredients one should
incorporate into an interest rate model in order to reproduce the observed statistical properties.  We then proceed to give our framework
a mathematical formulation in terms of a {\it stochastic evolution equation} in a space of smooth functions in section \ref{ees}. In the case where only {\it local} deformations are allowed, this equation reduces to a stochastic partial
differential equation: we study a simple example of such an evolution equation
in section \label{spde} and show that, albeit its rudimentary structure,
it reproduces many properties of term structure deformations in a simple 
manner.

\section{Statistical properties of term structure deformations}\label{empirical}

As in any applied discipline, empirical observations should be the starting
point in the construction of stochastic models in finance.
Since our aim here is to model the dynamical behavior of the yield curve, we shall
begin by describing some  important empirical facts about term structure
deformations. 
The results outlined in this section we mainly refer to \cite{string}, \cite{risk} and
\cite{litterman}.

\begin{enumerate}
\item Smoothness in maturity: yield curves do not present highly irregular
profiles with respect to maturity. Of course one could argue that with 50 or 60 data points it is difficult to assess the smoothness of a curve; this property should be viewed more as a requirement of market operators.
A "jagged" yield curve would be considered as a peculiarity by any market operator.
This is reflected in the practice of obtaining implied yield curves by smoothing data points using splines.

\item Irregularity in time: The time evolution of individual
forward rates (with a fixed time to maturity) are very irregular. This should be contrasted with
the regularity of forward rates with respect to time-to-maturity
and reveals an {\it asymmetry} between the respective roles of the variables
$t$ and $\theta$.

\item Principal components: Principal component analysis of term structure deformations
indicates that {\it at least} two factors of uncertainty are needed to model 
term structure deformations.  In particular, forward rates of different maturities
are imperfectly correlated.
Empirical studies \cite{string,litterman} uncover the 
influence of
a {\it level} factor which corresponds to parallel shifts of the yield curve,
 a {\it steepness } factor which corresponds to opposite changes in
short and long term rates and a {\it curvature} factor 
which influences the curvature of the yield curve. More precisely, the third
principal component, when projected on forward rates of different maturities,
shows a large component at maturities around one year and small 
coefficients on the two extremities of the yield curve \cite{string}.

\item Humped term structure of volatility: Forward rates of different maturities
are not equally variable. Their variability, as measured for example by the standard
deviation of their daily variations, has a humped shape as a function of the maturity,
with a maximum at $\theta \simeq $ 1 year and decreases with maturity beyond
one year \cite{string}. This hump is always observed to be skewed towards
smaller maturities. Moreover, although the observation of a single  hump
is quite  common \cite{moraleda}, multiple humps are never observed in
the  volatility term structure.  

\item A multivariate process:   while  many previous models focused
exclusively on the short rate process, trying to represent it as a
Markov process, recent econometric studies seem to reject the Markov
hypothesis for an  interest rate of a given maturity, the short rate 
in particular, pointing out
to the interdependence between interest rates which calls for
a multivariate approach. 

\end{enumerate}

\section{Modeling strategy}\label{modeling}

What are the lessons to be drawn from these empirical observations?
We will now try to define some criteria which a model should try to respect
in order to give a  ``faithful" statistical representation of interest rate
fluctuations.

\subsection{Role of the short rate}

First, the actual dynamics (as opposed to the risk-neutral dynamics)
of the forward rate curve cannot be reduced to that of the short rate:
the statistical evidence points out to the necessity of taking into
account more degrees of freedom in order to represent in an adequate fashion
the complicated deformations of the term structure.
In particular, the imperfect correlation between maturities and the rich variety of
term structure deformations shows that a one factor
model is too rigid to describe yield curve dynamics.

Furthermore, in practice the value of the 
short rate is either fixed or at least
strongly influenced by an authority exterior to the market (Federal Reserve, central banks),
 through a mechanism
  different in nature from that which determines rates of higher maturities which
are negotiated on the market. The short rate can therefore be viewed as an
exogenous stochastic input  which then gives rise to a deformation of the term structure
as the market adjusts to its  variations.
It is therefore plausible from an economic point of view to model
separately the dynamics of the short rate.

Second, as shown by Ait Sahalia in a recent study \cite{aitsahalia}
the short rate and the long rate (or equivalently, the short rate and
the spread) can be reasonably described by a bivariate diffusion
such as the one considered by  \cite{brennan} or \cite{schaefer}.

\subsection{Sources of randomness}

Traditional term structure models such as \cite{vasicek,cir,hjm} 
define --implicitly or explicitly--
the random motion of an infinite number of forward rates as diffusions driven by a 
finite number of independent Brownian motions.  This choice may appear
surprising since it introduces a lot of constraints on the type of evolution one can ascribe
to each point of the forward rate curve and greatly reduces the dimensionality i.e. the number of degrees
of freedom of the model, such that the resulting model is not able to reproduce any more
the complex dynamics of the term structure. Multifactor models \footnote{For a concise
review of multifactor term structure models see \cite{duffie}.}
are usually justified by refering to the results of principal component analysis
of term structure fluctuations. As remarked above (Sec. \ref{empirical}),
it is often observed that the first three principal components explain
more than 95\% 
of the observed variance of forward rates, suggesting that a three factor model
would be sufficient.
However, one should note that the quantities of interest when dealing with
the term structure of interest rates are not the first two moments of the forward 
rates but typically involve expectations of non-linear functions of the forward rate curve:
caps and floors are typical examples from this point of view.
Hence, although a multifactor model might explain the variance of the forward rate itself,
the same model may not be able to explain  correctly the variability of 
portfolio positions involving non-linear combinations of the same forward rates.
In other words, a principal component whose associated eigenvalue is small
may have a non-negligible effect on the fluctuations of a non-linear function of
 forward rates. This question is especially relevant when calculating quantiles and
Value-at-Risk measures. 

In a multifactor model with $k$ sources of randomness, one can use {\it any} 
$k+1$
instruments to hedge a given risky payoff. However, this is not what
traders do in real markets: a given interest-rate contingent payoff
is hedged with bonds of the same maturity.\footnote{Unless, of course, liquidity
considerations impose the trader to do otherwise.}
These practices reflect the existence of a risk specific to instruments
of a given maturity. 
The representation  of a maturity-specific risk means that, in a continuous-maturity limit one must also allow the number of sources of randomness to grow
with the number of maturities; otherwise one loses the localization in maturity
of the source of randomness in the model.
This point is  discussed
in more detail in Sec. \ref{local}.

\subsection{The Markov property}

An important ingredient for the tractability of a model is its 
Markovian character. Non-Markov processes are difficult to simulate and even harder to
manipulate analytically. Of course, any process can be transformed into
a Markov process if it is imbedded into a  space of sufficiently high dimension;
this amounts to injecting a sufficient number of "state variables" into the model. These state variables may or may not be observable quantities; for example
one such state variable may be the short rate itself but another one could be
an economic variable whose value is not deducible from knowledge of the forward rate curve. If the state variables are not directly observed, they are obtainable in principle
from the observed interest rates by a filtering process\cite{lacoste}. 
Nevertheless the presence of unobserved state variables makes the
model more difficult to handle both in terms of interpretation and 
statistical estimation. 
This drawback has motivated the development of so-called { affine curve models} models where one imposes
that the state variables be affine functions of the observed yield curve
\cite{duffie}.  While the affine hypothesis is not necessarily 
realistic  from an empirical point of view, it has the property
of directly relating state variables to the observed term structure.
We will try to conserve  this desirable feature  in our model.

\subsection{Continuity of term structure deformations}\label{continu}

\begin{center}
\begin{figure}
\centerline{\hbox{\epsfig{figure=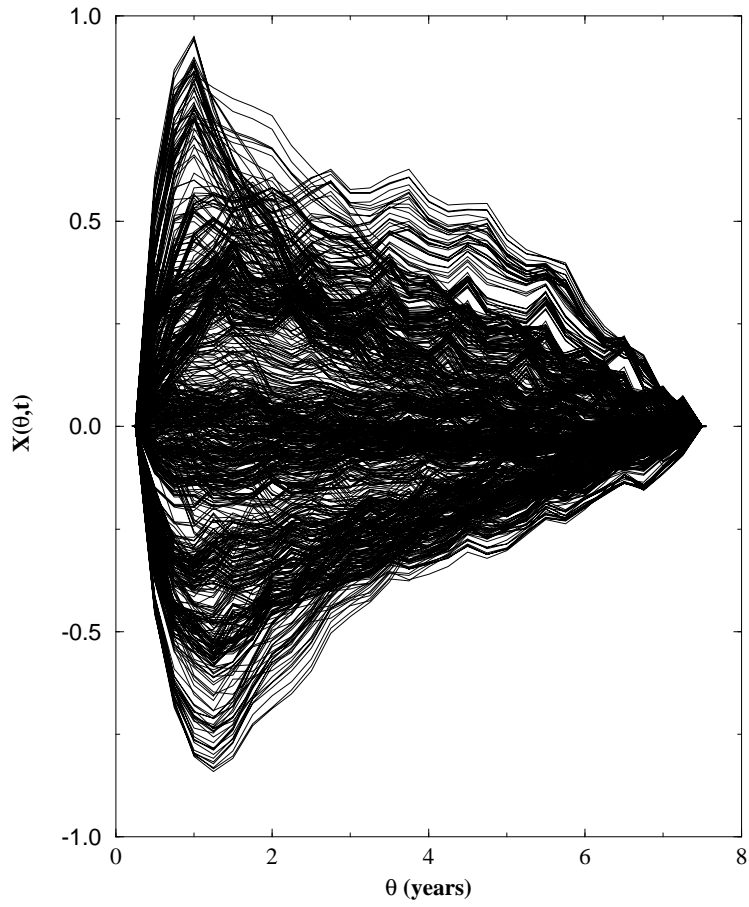,width=8cm}}}
\vskip 0.8cm
\label{fig3}
\caption{Various configurations adopted by the function $X_t(\theta)$ defined 
in Eq. \ref{parametrisation} as obtained from Eurodollar future contracts (1992-7). This figure may be
seen as a visualization of the stationary density  of the process $X_t$.}
\end{figure}
\end{center}

Another feature of term structure movements is that, as a curve, the forward rate curve
displays a continuous deformation: configurations of the forward rate curve at
dates not too far from each other tend to be similar. 
An animation movie of the successive positions of the forward rate curve 
displays
a continuous movement where the observer
can follow a given point on the curve\footnote{Redisplay Fig.\ref{fig3} in a postscript version of this article and watch!}.
This continuity  of deformations should be properly defined in mathematical
terms and accounted for in a term structure model.

\subsection{Smoothness in maturity}

As already noted above, most applications require the yield curve to
have some degree of smoothness e.g. differentiability with respect to
the maturity $\theta$. This is not only a purely mathematical requirement
but is reflected in market practices of hedging and arbitrage on fixed income
instruments. Market practitioners tend to hedge an interest rate risk of a given
maturity with instruments of the same maturity or close to it. This important 
observation means that the maturity is not simply a way of indexing the family
of forward rates:  market operators expect forward rates whose maturities
are close to behave similarly. Moreover, the model should account for the
observation that the volatility term structure displays a hump but that
multiple humps are never observed. Such an effect could be taken into account
by the presence of a term which couples together each forward rate with its
neighboring points on the yield curve. We will elaborate more on this aspect in
Sec. \ref{edps}. 

\section{A stochastic evolution equation for term structure deformations} \label{ees}

Based on the above considerations, 
we will now proceed to describe the deformations of the term structure in 
 mathematical form by means of a {\it stochastic evolution equation}, 
translating each of the criteria outlined above into their mathematical
equivalents.

\subsection{Definitions and notations}

We will parametrize the evolution of the term structure of interest
rates by the instantaneous forward rate curve (FRC), denoted by $f_t(\theta)$
where the subscript $t$ denotes time and $\theta \in [\theta_{\min},\theta_{\max}]$
 the  time to maturity.  Note that some authors (e.g. \cite{hjm}) specify the forward rate curve
as a function of the maturity {\it date} $T$; our parametrisation is related to the HJM parametrization in Eq.(\ref{hjm.eq}) by 
\ba
f_{\sc HJM}(t,T) &=& f_t(\theta=T-t)
\ea
As remarked in \cite{musiela}, this parametrization has the advantage that the 
forward rate curve process $f_t$ will belong to the same function  space
(a space of continuous curves defined on $[\theta_{\min}, \theta_{\max}]$)
when $t$ varies, which is not the case of  the process $f_{\sc HJM}(t, .)$
whose domain of definition $[t,T]$ shrinks with time.
Here  $\theta_{\min}$ is the shortest maturity available on the market
and
$\theta_{\max}$ the longest. 
$r(t) = f_t(\theta_{\min})$ will be called the {\it 
short rate}, $l(t) = f_t(\theta_{\max})$ the long rate.
The quantity $s(t) = l(t) - r(t)$ is the {\it spread}.

In most interest rate 
models $\theta_{\min}$ is taken to be $0$ and $\theta_{\max}=+\infty$ but this is
not necessarily the best choice nor even realistic.
First, it is obvious that in empirical applications maturities have a
finite span and $\theta_{\max}$ will be  typically 30 years or less
depending on the applications
considered. Second, the finiteness of $\theta_{\max}$ avoids some embarassing
mathematical problems related to the $\theta\to\infty$ limit \cite{dybvig,elk}
which are not necessarily meaningful from an economic point of view.
More importantly, we shall see that the $\theta_{\max}=+\infty$
limit is not ``innocent" : setting   $\theta_{\max}$ to a large but
finite value can be {\it qualitatively} different from taking it to be infinite.

\subsection{Decomposition of forward rate movements}

As mentioned above, given the particular nature of the short rate and the 
well known role of the short rate and the spread as two principal factors,
we first proceed to ``factor" them out of the model and parametrize
the term structure as follows:
\ba \label{parametrisation}
f_t(\theta) &=& r(t) + s(t) [ Y(\theta) + X_t(\theta) ]
\ea
where $Y$ is a deterministic {\it shape} function defining the average profile of
the term structure and $X_t(\theta)$ an adapted process decribing the
random deviations of the term structure from its long term average
shape. 
With no loss of generality we require:
\ba
Y(\theta_{\min}) = 0& &Y(\theta_{\max}) = 1
\ea
which results in
\ba
X_t(\theta_{\min}) = 0& &X_t(\theta_{\max}) = 0
\ea
The process $X_t(\theta)$ then describes the fluctuations of
a random curve with fixed endpoints. We will thus call 
$X_t$ the deformation of the term structure at time $t$.
Factoring out the fluctuations of the first two principal components
then means modeling separately the process $(r(t),s(t))$ and the 
deformation process $(X_t)_{t\geq 0}$. 

In a Gaussian framework,  the uncorrelatedness of the principal components would entail their independence. In particular the first two principal components
(which are roughly the spread and the short rate) would be independent from
the deformation process $X_t$.
We will use this assumption as a working hypothesis:

\proclaim Assumption. The deformation process $X_t$ is
independent from the short rate $r(t)$ and the spread $s(t)$.

\subsection{The short rate and the spread: a bivariate Markov process}

As in \cite{brennan}, one can consider that the short rate and the long rate
(or equivalently, the short rate and the spread, see \cite{schaefer}) are well described by
a bivariate diffusion process:

\ba\label{endpoints}
dr_t &=& \mu_1(r_t,s_t) dt + \sigma_{1,1}(r_t,s_t) dW_t^1 + \sigma_{1,2}(r_t,s_t) dW_t^2\\\
ds_t &=& \mu_2(r_t,s_t) dt + \sigma_{2,1}(r_t,s_t) dW_t^1 + \sigma_{2,2}(r_t,s_t) dW_t^2\
\ea
where $W^1,W^2$ are two independent Wiener processes.
This formulation is empirically motivated by the econometric studies
refered to above \cite{aitsahalia} in which it was shown that 
the hypothesis of a bivariate diffusion for $(r_t, s_t)$
is not rejected by non-parametric tests while  it is rejected for $r_t$
taken individually. 

However,  the  only hypothesis we need here is the jointly Markovian
character of $(r_t, s_t)$; ,for example, the noise source  in
Eq.(\ref{endpoints})   could be replaced with a non-Gaussian
L\'evy process without modifying what follows.

\subsection{Term structure deformations as Markovian curves}

We are now left with the deformation process $(X_t)_{t\geq 0}$ to model.
The first requirement we impose on $X_t$ is its smoothness in maturity:
at a given time $t$, $X_t$ is a function defined on $[\theta_{\min},\theta_{\max}]$ 
determined by the forward term structure which, as remarked above,  
is a ``smooth" function of the time to maturity $\theta$. 
$X_t$ should therefore belong to a suitable space $H$ of smooth functions 
which  will then be the state space of our model. 
In view of interpreting our results in terms of principal component analysis,  
we would like the state space $H$ to have some Hilbert-like structure in order to 
define orthogonal projections of  $X_t$ onto a suitable basis of $H$.

The second requirement we impose is that $X_t$ be a Markov process in $H$.
This property, as remarked in \cite{musiela},
 is already verified in the Heath-Jarrow-Morton \cite{hjm} framework
for the forward curve process $f_t$. Here we require slightly more, namely
that the Markovian character respect the factorial decomposition (\ref{parametrisation}). 
That is, we require the endpoints $ ( r(t), s(t) )$ and the deformation $X_t$
to be separately Markovian.

In a functional space, there are of course a wide variety of Markov processes.
It is the hypothesis of continuity in time of the deformation process (see (\ref{continu})
which enables to single out, the only class  among all Markov processes having this property,
namely {\it diffusions}. More precisely, stating that $X_t$ is a $H$-valued
diffusion process
means that there exist a drift functional $b$ and a volatility functional $\sigma$, defined on $H$, such that the evolution of $X_t$ is given by a stochastic
differential equation in $H$ (written here in Ito notation):

\ba
dX_t &=& \mu(X_t) dt + \sigma(X_t) d{\cal B}_t \label{diffusion}
\ea
where ${\cal B}_t$ is an appropriate generalization of Brownian motion taking values in $H$.
Here $\mu$ and $\sigma$ are allowed to depend on the contemporaneous term
structure i.e it can be a function of the whole curve $X_t(\theta), \theta\in[0,\theta^*]$.

Formally, Eq.(\ref{diffusion}) is a stochastic differential equation
in an (infinite-dimensional) functional space $H$. 
In order to give a proper meaning to Eq.(\ref{diffusion}), one should
start by specifying  the nature of the random noise source ${\cal B}_t$ such that the stochastic
integral implicit in Eq.(\ref{diffusion}) can be properly defined.
There are several ways to define a generalization of the Wiener process
and a stochastic integral in an infinite dimensional space. The relation 
between these different  constructions was clarified by Yor \cite{yor} 
who showed that the natural setting for constructing infinite dimensional 
diffusions is a Hilbert space. 
Given  a (separable) Hilbert space $H$  (for example
$H =L^2([\theta_{\min},\theta_{\max}],\nu)$ for some measure $\nu$),
one can define
 a {\it cylindrical Brownian motion} on $H$ as
 a family 
$({\cal B}_t)_{t\geq 0}$ of random linear functionals ${\cal B}_t:H\to R$ satisfying:
\begin{enumerate}
\item $\forall \phi \in H, {\cal B}_0(\phi) = 0$
\item
$\forall \phi \in H, {\cal B}_t(\phi)$ is an ${\cal F}_t$ -- adapted scalar stochastic process.
\item
$\forall \phi \in H-\{0\}, \frac{{\cal B}_t(\phi)}{|\phi|}$ is a one-dimensional
Brownian motion.
\end{enumerate}
In particular, if one takes any orthonormal basis $(e_n)$ in $H$ then its image
 $({\cal B}_t(e_n))_{t\geq 0}$
form a sequence of independent standard Wiener processes in {\bf R}.
This property is useful for building finite-dimensional approximations.

A suitable choice of state space verifying such requirements is a Sobolev space $H^s$, 
namely the space of functions $g\in L^2([\theta_{\min},\theta_{\max}],\nu)$ such that the s-th derivative $g^{(s)}$ is also in $L^2([\theta_{\min},\theta_{\max}],\nu)$, for some measure $\nu$ on $[\theta_{\min},\theta_{\max}]$. Given that the derivative 
$\partial f/\partial\theta$ is assumed to exist in many applications, 
we would like to require $s\geq 1$. See also the discussion of this point  
in \cite{douady}, where a similar choice is adopted in a slightly
different framework.

\section{Local deformations, stochastic PDEs and string models} \label{edps}

At this level of generality, not much can be said of the properties of the solutions
of Eq.(\ref{diffusion}). In this section we will show how the description
of term structure deformations through level, steepness and curvature of the
yield curve reduces Eq.(\ref{diffusion}) to a stochastic partial differential equation,
of which some simple examples are given.

\subsection{Market segmentation and local deformations}\label{local}

As mentioned before, the maturity $\theta$ is not simply a way to 
index different forward rates and instruments: the fact that
fixed income instruments are ordered by maturity is important for
market operators. For example, this is  reflected in the
hedging strategies of operators on the fixed income market:
to hedge an interest rate risk of maturity $\theta = $ 8 months,
an operator will tend to use bonds (or other fixed-income instruments)
of maturity close to 8 months: 6 months, 9 months. Although this strategy seems
quite sensible, it does not correspond to the picture given
by multifactor models: in a k-factor model, any $k+1$ instruments
can be used to hedge an interest rate contingent claim. For example in a two factor model
one could use in principle a 30 year bond, a 10 year bond and a 6 year bond
to hedge an instrument with maturity of two years! Needless to say, 
no sensible trader would follow such a strategy, which shows that
in practice the factors which explain 95\% of the variance are not enough
 to hedge 95\% of the risk of an instrument with a non-linear payoff:
this is precisely  our  principal motivations for introducing 
{\it maturity-specific }
sources of randomness i.e. one independent source of randomess per maturity.

The existence of maturity-specific risk naturally leads to a market
for such risk. Indeed, some  macroeconomic theories of interest rates
have considered the interest rate market as being segmented : for
example in a first approximation one can consider the market for
US Treasury bills, Treasury notes and Treasury bonds as being 3 separate markets where prices are fixed independently. In a continuous-maturity model
this would mean that the interest rate market is
partitioned into independently evolving markets involving instruments 
with maturity between $\theta$ and $\theta + d\theta$. 

However this is not strictly
true: as shown by principal component analysis,
long-term rates react to variations in the short rate
in a way that is not  explainable simply via parallel shifts and 
vertical dilations of the term structure. 
One way to conciliate the interdependence of rates of various
maturities with the  segmentation of markets across maturities is by
considering deformations of the term structure that are {\it local}
in maturity: a forward rate of maturity $\theta$ is more sensitive 
to variations 
of rates with maturity close to $\theta$.
We are not
dealing here with a strict segmentation of the market into separately
evolving markets but a ``soft'' segmentation which simply implies that
the market for each maturity adjusts itself to the variation in rates
of maturities immediately above and below it.
This means for example that, among all rates of maturity $\geq 1$ year,
the 1 year rate will have a higher sensitivity and react more quickly to
a variation in the short rate since it is closer in maturity.

\subsection{Level, steepness and curvature} 

In mathematical terms, the local deformation hypothesis
means that  the variation of $X_t(\theta)$ 
 will only depend on the behavior of $X_t(.)$ around $\theta$.
How can one parametrize the shape of the term structure around a  given
maturity $\theta$?
Given that $X_t:\theta \to X_t(\theta)$ is assumed to be a
 smooth curve, its local behavior around $\theta$ can be described by its first
few derivatives: $X_t(\theta),\quad {\partial_{\theta} X_t},\quad 
 {\partial^2_{\theta} X_t}$, ...

As noted before, empirical studies seem to identify the level of interest rates,
the steepness (slope) of the term structure and its curvature as three significant
parameters in the geometry of the yield curve \cite{litterman}.
In a market involving instruments of maturity between $\theta$ and $\theta +
d\theta$,
these three features are described by the level of rates, and the first
two derivatives with respect to $\theta$.
Combining the local deformation hypothesis  formulated in Sec.\ref{local}
with a local description of the term structure by level, steepness
and curvature one obtains that the drift and volatility of $X_t(\theta)$
can only depend on 
$X_t(\theta),\quad {\partial_{\theta} X_t},\quad 
 {\partial^2_{\theta} X_t}$. Therefore, Eq.(\ref{diffusion}) becomes
a second order stochastic partial differential equation:

\ba\label{spde.eq}
dX_t = [ \frac{\partial X_t}{\partial \theta} +
b(X_t(\theta),\frac{\partial X_t}{\partial \theta},\frac{\partial^2 X_t}{\partial \theta^2}) ]\quad dt +
\sigma(X_t(\theta),\frac{\partial X_t}{\partial \theta},\frac{\partial^2 X_t}{\partial \theta^2} ) dB_t(\theta)
\nonumber\\
\forall t \geq 0, X_t(\theta_{\max}) = X_t(\theta_{\min}) = 0\qquad
X_{t=0}(\theta) =
X_0(\theta) \qquad \ 
\ea
This equation is the mathematical expression of the fact that deformations are
local in maturity and that the deformation at maturity $\theta$ depends
 on the level, steepness and 
curvature of the term structure around $\theta$.

In the general case where $b$ and $\sigma$ are smooth  but nonlinear functions
of their arguments, Eq.(\ref{spde.eq}) is not easy to study: indeed, it is not
trivial to define properly what is meant by a solution of Eq.(\ref{spde.eq})
and even less to study their regularity. An approach to the fully non-linear case
using the notion of viscosity solution has been recently proposed
 for the case of a noise-source depending only on $t$ \cite{lions}.
In order to point out the differences with HJM-type models resulting from
the local deformation hypothesus, we shall consider the case 
of a forward rate dynamics such as (\ref{hjm.eq}) which is perturbated by  a term depending
on the curvature:
\ba\label{parabolic.eq}
dX_t = [ \frac{\partial X}{\partial \theta} +
b(t,\theta,X_t(\theta)) + \frac{\kappa}{2}\frac{\partial^2 X}{\partial \theta^2} ]\quad dt +
\sigma(t,\theta,X_t(\theta)) dB_t(\theta)
\nonumber\\
\forall t \geq 0, X_t(\theta_{\min}) = X_t(\theta_{\max}) = 0\qquad X_{t=0}(\theta) = X_0(\theta)
\ea
Properties of stochastic PDEs such as Eq.(\ref{parabolic.eq})
have been studied by Pardoux \cite{pardoux} and collaborators.
Properties of term structure deformations described by Eq.(\ref{parabolic.eq})
are studied in \cite{cont}. In the following section we study the simplest
case where volatility is constant; surprizingly, we will show that this
simple
case already presents many of the desirable features enumerated in Sec.\ref{modeling}.

\subsection{The linear parabolic case}\label{parabolic.sec}

In order to illustrate what are the type of dynamics implied by Eq.(\ref{parabolic.eq})
for term structure deformations, we will now study the simplest example of the above equations
which incorporates the influence of local steepness and curvature, namely the case where
$\sigma$ is independent of $X_t$. For the sake
of simplicity we will deal here with the constant volatility case
but all the results below remain valid in
the case of an arbitrary deterministic function of time $t$ (for details
see \cite{cont}).
The case of constant volatility leads us to the following stochastic partial differential equation\footnote{Up to the first derivative in $\theta$, this closely resembles what is known
as the stochastic heat equation in the PDE literature.}
:
\ba\label{parabolic}
\frac{\partial X}{\partial t} &=& [ \frac{\partial X}{\partial \theta}
+ \frac{\kappa}{2} \frac{\partial^2 X}{\partial \theta^2}] dt +
\sigma_0 dB_t(\theta)\\
\forall t \geq 0,& & X_t(\theta_{\min}) = X_t(\theta_{\max}) = 0\\
\forall \theta \in  [\theta_{\min},\theta_{\max}],& &  X_{t=0}(\theta) = X_0(\theta)
\ea

\subsection{Eigenmodes and principal components}

\begin{figure}
\epsfig{file=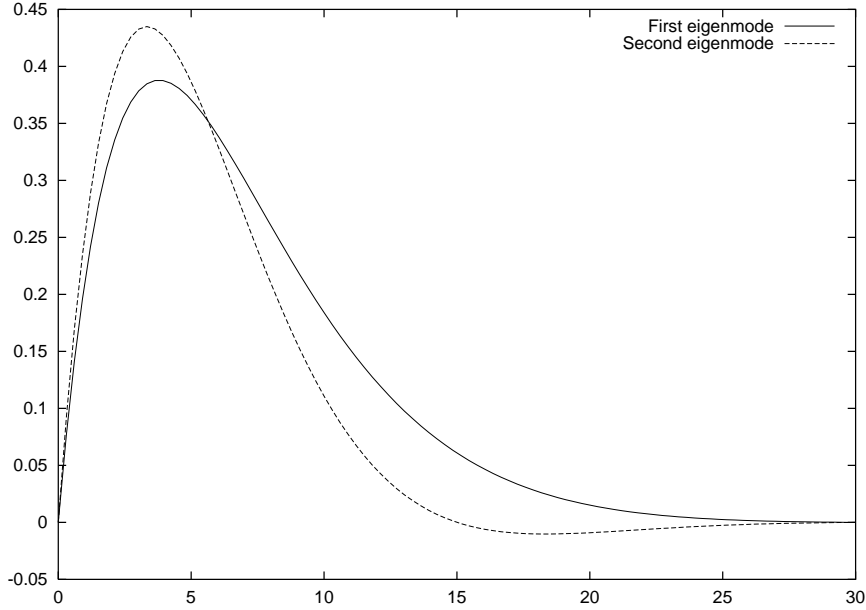,angle=270,width=12cm} \label{eigenfig}
\caption{First two eigenmodes of the operator A, with K=4 and $\theta^*=30$ years. Note the maxima situated at short maturities.}
\end{figure}

Let $\theta^{*} = \theta_{\max} - \theta_{\min}$  be the maturity span of
the observed forward rate curve. By translating the maturity variable
one can assume $\theta_{\min} =0$ without loss of generality in what follows.
We consider as state space for our solutions the Hilbert space
$H$ of real-valued functions defined on $[0,\theta^*]$ with the scalar product:

\ba
<f,g> &=& \int_{0}^{\theta^*} d\theta\quad \exp({\frac{2\theta}{\kappa}}) 
f(\theta) g(\theta) 
\ea
The subscript $H$ in $<.,.>_H$ will be omitted in most of this section.
Let $A$ be the operator in $H$ defined by:
\ba
A.u &=& \frac{\partial u}{\partial \theta}
+ \frac{\kappa}{2} \frac{\partial^2 u}{\partial \theta^2}
\ea
It is not difficult to show that
$A$ has a discrete spectrum, with eigenvalues and eigenfunctions given by:
\ba
A.e_n &=& -\lambda_n e_n\\
\lambda_n &=& 
\frac{1}{2\kappa} ({1 + \frac{n^2\pi^2\kappa^2}{ {\theta^*}^2 }})
\\
e_n(\theta) &=& \sqrt{\frac{2}{\theta^*}} \sin(\frac{n\theta\pi}{\theta^*}) \exp(-\frac{\theta}{\kappa}) \label{eigenfunction}
\ea
where $n$ takes all integer values $\geq 1$. Here the eigenfunctions $e_n$ have been normalized such that $(e_n)_{n\geq 1}$
is an orthonormal basis of $H$.
\ba
<e_n,e_m>&=&\quad \delta_{n m}
\ea
The functions (eigenmodes) $e_n(\theta)$ play the role of the principal components for the deformation process $X_t$. That is, if we perform a principal component analysis on a realization of the process $X_t$, for a large enough sample
the empirical principal components would reproduce the eigenmodes $e_n(\theta)$.
The first two of these eigenmodes are shown in Fig.(\ref{eigenfig}).
The role of the exponential term in Eq.(\ref{eigenfunction}) is clearly visible: the eigenfunctions become ``skewed'' towards shorter maturities and only
a single hump, whose position is determined by the value of $\kappa$, is visible. Recall that this exponential term stems simply from the fact that
we are parametrizing the forward rate process by time to maturity $\theta$
instead of maturity date $T$ \cite{musiela}. In particular,
in contrast with multifactor models \cite{moraleda},
there is no need to use a complicated volatility structure $\sigma(t,\theta)$ 
to obtain a volatility hump.  The position of the hump
gives a (first) simple method for calibrating the value of $\kappa$ to empirical observations.

The Green function (propagator) associated to the operator $A$
may then be expressed in terms of an eigenmode expansion:
\ba
G(t,x,y) &=& \sum_{n\geq 1}\exp(-\lambda_n t) e_n(x) e_n(y)
\ea
and Eq.(\ref{parabolic}) can be properly defined in  the following integral
form:
\ba
X_t(\theta) &=&\int_{0}^{\theta^*} G(t,\theta,y) X_0(y) dy +
\int_{0}^{t} ds \int_{0}^{\theta^*} G(t-s,\theta,y) \sigma_0 d{\cal B}_s(y) \label{intparabolic}
\ea 
Let $X_t(\theta)$ be the solution of Eq.(\ref{intparabolic}).
Define the coordinates of the solution in the eigenvector basis as:
\ba
x_n(t) = <X_t,e_n> &=& \int_{0}^{\theta^*} \quad d\theta\quad X_t(\theta) e_n(\theta) \quad \exp({\frac{2\theta}{\kappa}}) 
\ea
The coefficients $x_n(t)$ therefore represents the projection of the deformation process $X_t$ on the $n$-th principal component (eigenmode).
For each $n$, $x_n(t)$ is then a solution of a linear stochastic differential equation:
\ba
dx_n(t)&=& -\lambda_n x_n(t) dt  + \sigma_0 dW^n_t
\ea
where $W^n_t = {\cal B}_t(e_n)$ are independent  standard Wiener processes.
The $x_n(t)$ therefore consitute a sequence of independent Ornstein-Uhlenbeck processes:
\ba \label{decay}
x_n(t) &=& e^{-\lambda_n t} <e_n,X_0> + \int_{0}^{t} ds\quad e^{-\lambda_n (t-s)} \sigma_0 dW^n_s
\ea
This last equation has an interesting interpretation. Remember that $x_n(t)$
the projection of the deformation process $X_t$ on the $n$-th principal component. Eq.(\ref{decay}) expresses $x_n(t)$ as the sum of two components, the first one being the contribution of the initial term structure to $x_n$ and the second one its stationary value. 
Eq.(\ref{decay})  may then be interpreted by stating that the 
forward rate curve ``forgets'' the contribution of the $n$-th principal component to the initial term structure at an exponential rate with characteristic time
\ba
\tau_n = \frac{1}{\lambda_n} = \frac{2\kappa}{{1 + \frac{n^2\pi^2\kappa^2}{ {\theta^*}^2 }} } 
\ea
Therefore, a perturbation of the initial term structure due to the $n$-th principal component will disappear or be smoothed out after a typical time $\tau_n$
which decreases with $n$: ``singular'' perturbations die out more quickly than
smoother ones. This shows the important relation between
the smoothing property of the deformation operator $A$ and the decay of its
eigenvalues: an operator with a quickly decaying spectrum will
guarantee  a fast decay (in time) of the singularities appearing
in the term structure and restore smoothness in maturity.  This gives a second
interpretation of the parameter $\kappa$: in addition to determining the position of the volatility hump, it also determines the decay rate of perturbations
of the term structure. This interpretation gives a second, independent method
for calibrating the model parameters to empirical data.
 One can easily imagine more general models where these two roles
of $\kappa$ can be attributed to two separately calibrated parameters \cite{cont}.

The Ornstein-Uhlenbeck process is the process used to represent the short rate
in \cite{vasicek}: it possesses the fundamental property of {\it mean reversion}
which has made it a popular model in interest-rate modeling. In our case,
this mean reversion is observed in  the principal components of the yield curve:
while individual forward rates may have a non-stationary and irregular behavior
the yield curve as a whole will converge to a stationary state with a mean-reverting behavior.

Given the explicit form of $\lambda_n$, it is easy to show that the series
\ba
E[X_t(\theta)] &=& \sum_{n\geq 1} E[x_n(t)] e_n(\theta)\\
Var[X_t(\theta)] &=& \sum_{n\geq 1} Var[x_n(t)] e_n(\theta)^2
\ea
are absolutely convergent for all $(t,\theta)$ and the sum
\ba
X_t(\theta) &=& \sum_{n\geq 1} x_n(t) e_n(\theta)
\ea
defines a unique Gaussian random field  
with mean and variance given by:
\ba\label{variance}
E[X_t(\theta)]&=& \sum_{n\geq 1} e^{-\lambda_n t} <X_0, e_n> e_n(\theta)\\
Var [X_t(\theta) ]&=& \sigma_0^2 \sum_{n\geq 1} \frac{1-\exp(-2\lambda_n t)}{2\lambda_n} e_n(\theta)^2
\ea

\subsection{Average term structure and mean reversion}

Under the 	above assumptions, one can calculate the average shape of the  term structure
of forward rates from (Eq. \ref{parametrisation}).
\ba
E[f(t,\theta)]= E[r(t)] + E[s(t)] Y(\theta)
\ea

The shape function $Y(\theta)$ can therefore be chosen in order to
reproduce the average term structure. In \cite{string} it was found that
the function
\ba
Y(\theta)&=&\sqrt{\frac{\theta}{\theta*}}
\ea
gives a good fit of the average shape of the Eurodollar term
structure for maturities ranging from 3 months to 10 years.
However, the precise analytic form of  the shape function $Y$
does not affect the results above. How does the yield curve fluctuate 
around its average shape? It is easily seen from Eq.(\ref{variance})
that the process $X_t$ converges to a Gaussian random field $X_\infty$
with mean zero and covariance:
\ba
Cov(X_t(\theta),X_{t'}(\theta')) &=& \sigma_0^2 \sum_{n\geq 1} \frac{e_n(\theta)
e_n(\theta') e^{-\lambda_n(t-t')}}{2 \lambda_n}
\ea

In terms of term structure movements, stationarity of term structure deformations
implies a mean-reverting behavior of forward rates. But a model such as the above 
asserts more: it enables to calculate (therefore calibrate, if one is interested in
such quantities) the probability
of a given yield curve deformation.

\subsection{Random strings: parabolic vs. hyperbolic formulation}

In a recent work \cite{santaclaus} it has been  proposed to consider stochastic partial differential equations of {\it hyperbolic} type to describe
the evolution of the forward rate curve. 
Such an equation differs from the above one through the presence of a second-order time derivative which dominates the dynamics. 
The model in \cite{santaclaus} is formulated
in terms of the forward rate process itself (in the spirit of \cite{hjm})
 and not in terms of the deformation
 process $X_t$, by examining the effect of 
inserting a  second-order time derivative in the equations of Sec. \ref{edps}.
Let us therefore
 consider the general case where the evolution equation contains 
both a propagation term and a diffusion term:

\ba\label{general}
f \frac{\partial^2 X}{\partial t^2}
+ \frac{\partial X}{\partial t} &=& \frac{\partial X}{\partial \theta}
+ \frac{\kappa}{2} \frac{\partial^2 X}{\partial \theta^2} +
\sigma_0 dB_t(\theta)\\
\forall t \geq 0,& & X_t(0) = X_t(\theta^*) = 0\\
X_{t=0}(\theta) &=& X_0(\theta) \frac{\qquad \partial X_{t=0}}{\partial t}= Y_0
\ea
The above equation is analogous to that of a vibrating elastic string,
hence the name of ``string models'' given to such descriptions of term
structure movements.
The case $f=0$ is the one studied in Sec.\ref{edps}; the case $f\to\infty$ 
(the other parameters being appropriately rescaled) is the stochastic wave equation
\cite{dz}, the ``space" variable being $\theta +t$. Note that
the stochastic PDE used in \cite{santaclaus} is formulated as a PDE
perturbated by  a two-parameter
("space-time") noise (called a ``stochastic string shock")
 while our Eq.(\ref{parabolic.eq}) or the general case   Eq.(\ref{spde.eq}) was presented  above
as an evolution equation for a curve in some function space. In the case of a parabolic
SPDE, where only the first derivative with respect to time is involved,
the two approaches are equivalent for a two-parameter process. Choosing one
approach or the other then amounts to viewing  the solution of a stochastic PDE either as a 
random field or as a stochastic process in a function space. Implicit in this choice
is whether the object of interest for modeling purposes is an individual interest rate
or the deformation of a multivariate object, namely the term structure.
The parabolic equation in Sec.\ref{parabolic.sec} has the merit of emphasizing the asymmetric roles of the variables
$\theta$ and $t$, an empirically desirable feature which is not  present,
 as we shall see below, in \cite{santaclaus}.

 Manifestly,
the operator on the right hand side of Eq.(\ref{general}) is the same
as in Eq.(\ref{parabolic}). This means that the deformation eigenmodes (the eigenfunctions of $A$) will remain the same as
in Eq.(\ref{parabolic}) studied above
but the projection of the process $X_t$ on each of them 
will be different. For example,
a stationary solution of Eq. (\ref{general}) will not give the same weight to 
the eigenmodes as and therefore the results of a Principal Component Analysis of Eq.(\ref{general}) will differ from that of Eq.(\ref{parabolic}) in terms of the eigenvalues.

The representation of the equation in terms
of its projections on the eigenmode basis $e_n$ gives as above a stochastic equation for the scalar process
$x_n(t)$:

\ba
f \frac{d^2 y_n}{dt^2}
+ \frac{dy_n}{dt} &=& -\lambda_n y_n(t) + \sigma_0 \dot{W}^n_t
\ea
This formal second-order stochastic differential equation is interpreted in the usual way, as follows.
Consider the Green function $U_n(t)$ of the operator $f \frac{\partial^2 u}{\partial t^2}
+ \frac{\partial u}{\partial t} + \lambda_n u$, given by:
\ba
U_n(t) &=& \frac{1}{f \sqrt{1- 4\lambda_n f}} [ e^{r_{1,n}t} - e^{r_{2,n} t}] 1_{t>0}\\
\ea
where $r_{1,n}$ and $r_{2,n}$ being the roots of the associated characteristic equation:
\ba \label{carac}
f r^2 + r + \lambda_n = 0.
\ea 
The process $x_n(t)$ is then given by the stochastic convolution integral:
\ba
y_n(t) &=& a_n e^{-r_{1,n} t} + b_n e^{-r_{2,n} t} + \int U_n(t-s) dW^n_s  
\ea
which is well-defined since the integrator is square-integrable in $s$.
Here $a_n$ and $b_n$ are defined by the initial term structure. Depending on
the values of $f$ and $\kappa$, two scenarios are possible:
\begin{enumerate}
\item Oscillating initial conditions: if 
\ba\label{flarge}
f &>& \frac{\kappa}{2 (1+ \frac{\pi^2\kappa^2}{{\theta^*}^2})} 
\ea
then for all $n\geq 1$, Eq.\ref{carac} has two complex conjugate roots
given by:
\ba
r_{1,n} &=& \frac{i \sqrt{4\lambda_n f -1} - 1}{2f} = -\frac{1}{2f} + i \omega_n\label{root1}\\
r_{2,n} &=&  \frac{- i\sqrt{4\lambda_n f - 1} - 1}{2f} = -\frac{1}{2f} - i \omega_n
\ea
The real part gives an exponential damping of the initial conditions which
characterizes the mean reverting behavior of $X_t$ as in the parabolic case.
First remark  that, unlike the parabolic case where ``bumpy" principal components
which contribute the most to non-smoothness in maturity decay more quickly,
here all principal components decay with the same speed i.e. a mean reversion time
of $2f$.
Recall that $\kappa$ still determines the position of the volatility hump
 so $\kappa\simeq$ 1 year.
So  (\ref{flarge}) implies that $f > 6$ months.
The mean reversion time of the whole curve is thus around  a year.
However, a new  phenomenon appears: the principal components do not simply revert
to their mean but {\it oscillate} around their mean with a frequency $\omega_n/ 2\pi$
which increases with $n$:
\ba
x_n(t) &=& A_n e^{-t/2f} \cos (\omega_n t + \phi_n)
\ea
The phase $\phi_n$ and amplitude factor $A_n$  are determined by (two)
initial conditions (see below). 
The oscillation of the term structure around its mean is not necessarily
an undesirable feature of this model and indeed can be justified on economic grounds \cite{cont}.
But the slow mean reversion combined with increasingly faster oscillations
of the higher order principal components leads to non-smoothness in maturity
of the solutions of Eq.(\ref{general}) \cite{dz}: in fact one should expect the
cross-sections in time or maturity to have the same irregularity which,
as pointed out in Sec.\ref{modeling}, is not
a desirable feature for a term structure model.

\item Selective damping of principal components: if
\ba
f &<& \frac{\kappa}{2 (1+ \frac{\pi^2\kappa^2}{{\theta^*}^2})} 
\ea
then
$\exists N>1$ such that for $ n\leq N $ Eq.(\ref{carac}) has two real, negative roots
whereas for $ n > N $ the roots are complex conjugates with negative real parts.  The projections of the deformations process $X_t$ on the first $N$
eigenmodes will have  a mean reverting behavior as in the parabolic case\footnote{In fact the decay of the initial
condition is described in this case by the superposition of two decreasing exponentials with
time constants given by $r^{-1}_{1n}$ and $r^{-1}_{2n}$},
with a mean reversion time increasing with $n$.
For $n > N$, $x_n(t)$ will  have a damped oscillatory behavior, with a damping time $\tau=2f$  
independent of $n$  and an oscillation frequency increasing with $n$ as above.
\end{enumerate}

Another crucial difference between Eq.(\ref{general}) and Eq.(\ref{parabolic.eq})
is the nature of the initial conditions. In the case of the parabolic equation (\ref{parabolic.eq})
the problem has a well defined solution once the initial term structure is specified 
through $X_0$. This is not sufficient in the case of Eq.(\ref{general}):
one must also specify the derivative  with respect to time at $t=0$.
In the case of a vibrating string, this means specifying the initial position and the
initial velocity of each point of the string. For a model of the forward rate curve,
this can be inconvenient: while the initial term structure is the natural input
for the initial condition of a dynamic model, the time derivative of the forward rates
is not easily evaluated, especially given the irregularity in time of forward rate trajectories
which prevents such a model from being calibrated in a numerically stable manner \cite{cont}.

We therefore conclude that the 
question of including a second-order time derivative in Eq.(\ref{general}) is not simply a matter of taste: the presence of
a second derivative radically changes both the dynamic properties of the equation and the nature of the initial conditions needed to calibrate the model, in
an empirically undesirable fashion. Our analysis thus pleads for a description 
of yield curve deformations
through a parabolic rather than hyperbolic SPDE.

\section{Conclusion and perspectives}

We have presented a simple stochastic model for describing the 
fluctuations of the term structure of forward rates: the forward rate
curve is described as a random curve oscillating around its long term
average.
The model studied in Sec. \ref{edps} should be viewed as the simplest
example of the type of model presented in Sec. \ref{ees}. However this simple example has the benefit of emphasizing the role of the
second derivative with respect to maturity in the evolution of the term structure: indeed, as we have seen above, it is this second derivative which
tames the potentially infinite number of sources of randomness and maintains a regularity
in $\theta$ while allowing for independent shocks along maturities.
It also gives the correct  form for the principal components as well as a
qualitatively correct estimate for their associated eigenvalues. 
These results show the importance of the concept of local deformation
 explained in Sec.\ref{local}, of which
our equation is the simplest example.
The model in Sec.\ref{parabolic.sec} can be easily generalized to the
case where the volatility surface $\sigma(t,\theta)$
is an arbitrary deterministic function \cite{cont}. The introduction of
a non-linear drift $b$ depending on the level of interest rates -- as in Eq.(\ref{parabolic.eq}) --
is also possible. More general cases remain to be studied.

As mentioned in the introduction, our objective has been to obtain a faithful
continuous-time representation of the statistical properties of the forward rate curve.
What remains is to establish the link with the arbitrage pricing approach
and examine the constraints imposed by absence of arbitrage on models of the type
exhibited above. Previous work in this direction \cite{douady} indicates that
such an analysis requires a careful reconsideration of the class of  arbitrage strategies
one is willing to consider in the context of  a partially segmented market such
as the fixed-income market.

\end{document}